\documentclass[a4paper]{article}

\usepackage[english]{babel}
\usepackage[utf8]{inputenc}
\usepackage{amsmath}
\usepackage{graphicx}
\usepackage[colorinlistoftodos]{todonotes}

\title{A Mathematical Model of Cell Reprogramming due to Intermediate Differential Regulator's Regulations}

\author{Arnab Barua\\ Department of Physics, IIT Bombay\\Email:-155120003@iitb.ac.in}

\date{\today}

\begin{document}
\maketitle

\begin{abstract}
In this paper I have given a mathematical model of Cell reprogramming from a different contexts. Here I considered there is a delay in differential regulator rate equations due to intermediate regulator's regulations.  At first I gave some basic mathematical models by Ferell Jr.[2] of reprogramming and after that I gave mathematical model of cell reprogramming by Mithun Mitra[4]. In the last section I contributed a mathematical model of cell reprogramming from intermediate steps regulations and tried to find the critical point of pluripotent cell .
\end{abstract}

\section{Introduction}
Basic unit of biological organisms are cells. In mammals sperm fertilize the egg and make an embryonic stem cell, which is a single cell. A beautiful thing in biological system is that all the biological organisms are made from the single cell. We can find 200 types of cells in our body and their jobs are different. They have same DNA sequence. This phenomena is known as Cell Differentiation.Where a cell type changes another cell type. Cell signals play a major role in this context externally and internally.    So, the question is that containing same kind of information how will you get different kind of cell types in your body?Can we understand this biological process from dynamical systems point of view?
\\Another beautiful phenomena is known as Cell reprogramming where the information contained in cell is removed and due to external stimulus it moves into a different cell type. For this John Gurdon and Shinya Yamanaka got the Noble Prize in 2012. In this paper I studied the Cell differentiation and Cell Reprogramming phenomena from dynamical systems point of view.
\section{An Intuitive Picture and Biology of Cell Differentiation}
At first I will give an intuitive picture. So let you and your friend went to a library and borrowed some books. Say your favorite subject is Physics and your friend's favorite subject is History. When both of you read this book , your mind will definitely change.But all the books are in the library. So in biology language Cell differentiation phenomena is that in the same DNA sequence some genes are off and some are on (say 1011 is for eye cell and 1100 for Lip cell,1 means on 0 means off). It happens due cues which can be act internally or externally. DNA Methylation, Histone Acetylation, mRNA these play an important role in Cell Differentiation. 
\section{Mathematical Models for Cell Differentiation}
\label{sec:examples}

In 1957 C.H.Waddington[1] gave an intuitive picture where he considered Cells(as a balls)rolling down in a potential surface(as Hills and valleys). Where each valley is considered as a cell fate. Ferell Jr.[2] proposed some mathematical models which is quite similar with Waddington's Epigenetic Landscape. two models are 1)Cell fate induction model 2) Lateral inhibition Model.

\subsection{Cell Fate Model}

A cell can induce another cell to be differentiated and adopt a phenotype( or a group of cells can induce another group of cells to be differentiated and adopt a phenotype) by chemical signals. In this model he considered a single differential regulator of concentration x (Example- Transcription factors) which has a constant basal rate $\alpha$ , a maximum positive hill function for feedback rate of synthesis x is $\beta$ and a degradation rate $\gamma$.

The rate of concentration x is (Considering a high coefficient n = 5)
$$\frac{dx}{dt} =  \alpha + \beta\Bigg(\frac{x^5}{K^5+x^5}\Bigg) -\gamma{x}$$   
where K is the concentration of x when the feedback is half of the maximum.

The potential function is $\phi$ and $\frac{d\phi}{dx}$ is for the speed at which it's come to steady state.So, 
$$\phi = -\int\Big(\alpha+\beta(\frac{x^5}{K^5+x^5})-\gamma x\Big)dx$$
After Plotting this potential function $\phi$ vs. x,  there is two stable points at x=0 and x= 1.7 in Fig 1. and an unstable point at x=1. If you increase the value of $\alpha$ and $\beta$ , you will find that there is only one steady state(Fig 2.). So, it's a Saddle-node bifurcation.Also, it's different from Waddington's Epigenetic Landscape.
\\Basic difference with Waddington's Landscape is that in Cell Fate Model the valley is created or destroyed for the value of $\alpha$ and $\beta$. But in Waddington's Landscape the valley's are permanent but Cell will take the decision internally or externally at the critical point. 

Potential function($\phi$)(y axis) vs. x (x axis) plot.

$$\includegraphics[scale=0.5]{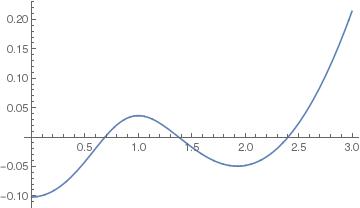}$$

Pic1($\alpha$=0,$\beta$=1,$\gamma$=0.5,K=1)
 $$\includegraphics[scale=0.5]{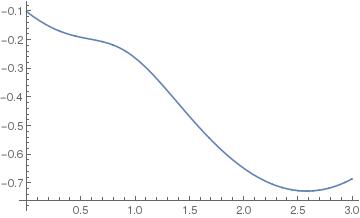}$$

Pic2($\alpha$=1,$\beta$=1,$\gamma$=0.5,K=1)

\subsection{Lateral Inhibition Model}

In this model he considered that two daughter cells m and n made from the mother cell x.The daughter cells m and n will mutually inhibit each other. The input signal of this model is interaction (I). After a critical value of interaction(I)=0.5 either m will win or n will win. He got Waddington's Landscape of Pitchfork Bifurcation. The mathematical model is

$$\frac{dm}{dt}=\alpha\Bigg(\frac{K^5}{K^5+(In)^5}\Bigg)-\beta m$$
$$\frac{dn}{dt}=\alpha\Bigg(\frac{K^5}{K^5+(Im)^5}\Bigg)-\beta n$$
where $\alpha$ is the coefficient of inhibition, $\beta$ is the coefficient of degradation and K is the concentration of x when the feedback is half of the maximum.
It's also depends on initial conditions of two daughter cells.

\section{Cell Reprogramming}
Cell reprogramming is a phenomena where  the epigenetic memory of a cell fate is removed by some stimulus and we can get another cell fate. Such an experiment was done by Nagy and Nagy[3]. They did this experiment and found differentiated fibroblast cells that were derived from induced pluripotency and also got four Yamanaka Factors under the control of Doxycycline drug.
\subsection{Mathematical model for Cell Reprogramming}
Mithun Mitra and their group[4] proposed a mathematical model of cellular reprogramming on the basis of Ferell's Cell Fate Model. They said that if we apply the Doxycycline for a certain time then we can get Pluripotent Cell from Somatic Cell. Also they considered the delay time for multiple chemical reactions,Cell shape and physical origins. They didn't take the delay term for degradation part.
The mathematical model is 
 $$\frac{dx}{dt} =  \alpha[d-t]+ \beta\Bigg(\frac{(x-p)^5}{K^5+(x-p)^5}\Bigg) -\gamma{x}$$
 Where $\alpha$ is the concentration of Doxycycline drug, $\beta$ is the coefficient of inhibition  and $\gamma$ is degradation coefficient.
p is the time of delay and d is the time up to which the Doxycycline drug is applied to the cell. First term here they considered the Heaviside step function for constant supply of drug upto certain time.
$$\includegraphics[scale=0.5]{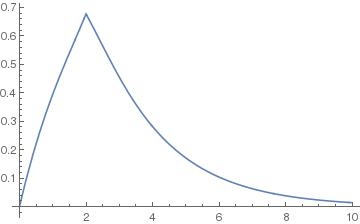}$$
Pic3($\alpha$=0.5,$\beta$=1,$\gamma$=0.5,K=1,p=0,d=2),Initial condition(x[0]=0)
they defined it's Somatic cell at x=0.Because in Ferell's Cell Fate Model there are two stable states at x=0 and x= 1.7
$$\includegraphics[scale=0.5]{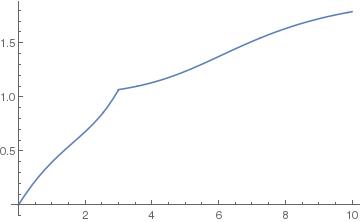}$$
Pic3($\alpha$=0.5,$\beta$=1,$\gamma$=0.5,K=1,p=0,d=3),Initial condition(x[0]=0)
It's the Pluripotent Cell.
But beautiful thing is that I find some meta stable states at x=4 for particular $\alpha$ ,$\beta$ and $\gamma$ shown in these pictures
$$\includegraphics[scale=0.5]{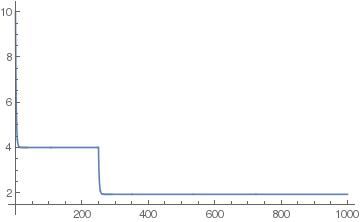}$$
($\alpha$=1,$\beta$=1,$\gamma$=0.5,K=1,p=0,d=250),Initial condition(x[0]=10)
$$\includegraphics[scale=0.5]{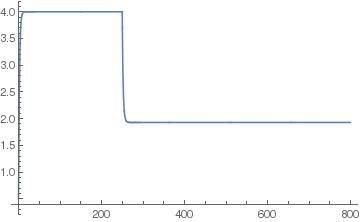}$$
($\alpha$=1,$\beta$=1,$\gamma$=0.5,K=1,p=0,d=250),Initial condition(x[0]=0.5)
\subsection{Considering Intermediate states in the regulation of cells and Doxycycline drug applied in one of them }
I proposed a mathematical model where I considered intermediate regulation in X(where some differential regulator of x will regulate differential regulator n and differential regulator m will regulate differential regulator  n .And the differential regulator n will regulate differential regulator x).
My mathematical model is
$$\frac{dx}{dt}=\alpha[d-t] + \beta\Bigg(\frac{m^5}{K^5+m^5}\Bigg) -\gamma{x}$$
$$\frac{dy}{dt}= \beta\Bigg(\frac{x^5}{K^5+x^5}\Bigg) -\gamma{y}$$
$$\frac{dz}{dt}= \beta\Bigg(\frac{y^5}{K^5+y^5}\Bigg) -\gamma{z}$$
$$\frac{dm}{dt}= \beta\Bigg(\frac{z^5}{K^5+z^5}\Bigg) -\gamma{m}$$
where $\alpha$ is the concentration of Doxycycline drug, $\beta$ is the coefficient of inhibition, $\gamma$ is degradation coefficient and K is the concentration of x when the feedback is half of the maximum.
Plotting x[t] Vs. time(t)-
$$\includegraphics[scale=0.5]{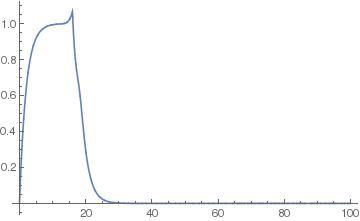}$$
at d = 16 (Somatic Cell)
$$\includegraphics[scale=0.5]{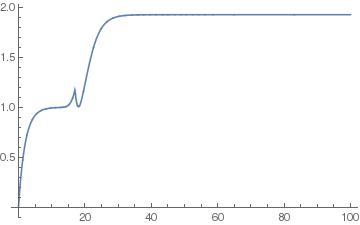}$$
at d=17 (Pluripotent Cell)
($\alpha$=0.5,$\beta$=1,$\gamma$=0.5,K=1),Initial condition(x[0]=0)
Similarly I find 'd' for Pluripotent Cell upto 5 intermediate cells.I got a linear relation between d and no.of intermediate cells(n). I plotted it (d in y axis and no. of intermediate states in x axis).
$$\includegraphics[scale=0.5]{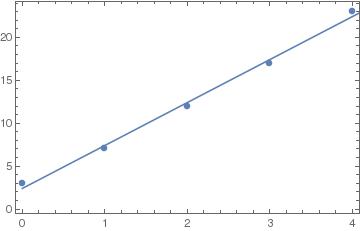}$$
and the deviation of the least fit straight line is (y axis)
$$\includegraphics[scale=0.5]{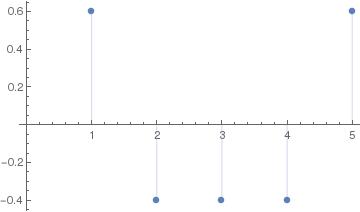}$$
The empirical equation of d
and n is= d = 2.4+5n
\subsection{Using Doxycycline in One cell what will be the effect
of that other cell
}
Here I want to see the effect in other cell due to Doxycycline in one cell. I considered the self regulation and mutual inhibition in this two cells. The mathematical model is
$$\frac{dx}{dt}=\alpha[d-t] +\beta\Bigg(\frac{K^5}{K^5+y^5}\Bigg)+ \delta\Bigg(\frac{x^5}{K^5+x^5}\Bigg) -\gamma{x}$$
$$\frac{dy}{dt}=\beta\Bigg(\frac{K^5}{K^5+x^5}\Bigg)+\delta\Bigg(\frac{y^5}{K^5+y^5}\Bigg) -\gamma{y}$$
Plot x[t] and y[t](Y axis) vs.time (x axis)
$$\includegraphics[scale=0.5]{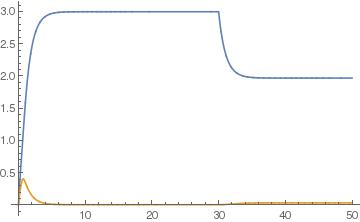}$$

(where  $\alpha$ is the concentration of Doxycycline drug, $\beta$ is the coefficient of inhibition,$\delta$ is the coefficient of self regulation  and $\gamma$ is degradation coefficient,\\$\alpha$=$\beta$=$\gamma$=1,,K=1,d=30,\\y[t]-Orange and x[t]-Blue,initial conditions y[0]=0,x[0]=0)
\subsection{Delay in Intermediate States}
Now,My motivation behind this mathematical model is that there is some biological steps due to chemical reactions which cause delay in regulation and  finding "Area 51" from this model. I considered the long delay in 2 intermediate states. I changed the previous mathematical model(Here I considered only 2 intermediate states only) with multiplying a small constant (I choose $\epsilon$= 0.1) to the differential equations corresponding y and z.Actually it's called forward feedback loop,which is more biologically relevant.
$$\frac{dx}{dt}=\alpha[d-t] + \beta\Big(\frac{z^5}{K^5+z^5}\Big) -\gamma{x}$$
$$\frac{dy}{dt}= \epsilon\Bigg(\beta\Big(\frac{x^5}{K^5+x^5}\Big) -\gamma{y}\Bigg)$$
$$\frac{dz}{dt}= \epsilon\Bigg(\beta\Big(\frac{y^5}{K^5+y^5}\Big) -\gamma{z}\Bigg)$$
where$\alpha$ is the concentration of Doxycycline drug, $\beta$ is the coefficient of inhibition ,K is the concentration of x when the feedback is half of the maximum and $\gamma$ is degradation coefficient.
\\x[t],y[t],z[t] along y axis and time along x axis
\\$\includegraphics[scale=0.5]{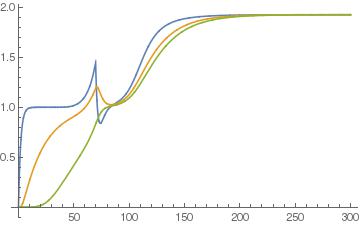}$
\\(d=69.67,$\alpha$=0.5,$\beta$=1,$\gamma$=0.5,Initial conditions x[0]=0,y[0]=0,z[0]=0,x is blue,y is orange,z is green).So I find some critical point at d = 69.5 after that the cell will go to Pluripotent cell.
\section{Summary and Acknowledgments}
In this paper I gave an mathematical model of Cell Reprogramming from intermediate differential regulator regulations, which is more biologically relevant.This work is supported by Prof. Sitabhra Sinha in IMSc Chennai. He always helped me in his busy time schedule.I am thankful to IMSc, Chennai for funding. I used Wolfram Mathematica software for generating plots.
\section{References}
[1] Epigenetic landscaping: Waddington's Use of Cell Fate Bifurcation Diagrams by Scott F. Gilbert.
[2]Bistability, Bifurcations, and Waddington's Epigenetic Landscape by James E. Ferrell Jr. 
[3]The mysteries of induced pluripotency: where will they lead? by Andras Nagy and Kristina Nagy.
[4]Delayed self-regulation and time-dependent chemical drive leads to novel states in epigenetic landscapes by Mitra MK, Taylor PR, Hutchison CJ, McLeish TC, Chakrabarti B.
\end{document}